\documentclass[seceq]{ptptex}

\usepackage{graphicx}
\usepackage{wrapft}

\title{
ANALYSIS OF THE RELAXATION PROCESS 
\\ 
USING NON-RELATIVISTIC KINETIC EQUATION %
}

\author{
Makoto \textsc{Takamoto}$^{1}$ 
and Shu-ichiro \textsc{Inutsuka}$^{2}$%
}

\inst{
$^1$Department of Physics, Kyoto University, Kyoto, 606-8502, Japan
\\
$^2$Department of Physics, Nagoya University, Nagoya, 464-8602, Japan
}

\abst{
We study the linearized kinetic equation of relaxation model 
which was proposed by Bhatnagar, Gross and Krook~\cite{bgk} (also called BGK model) 
and solve the dispersion relation.
Using the solution of the dispersion relation, 
we analyze the relaxation of the macroscopic mode and kinetic mode. 
Since BGK model is not based on the expansion in the mean free path in contrast to the Chapman-Enskog expansion,
the solution can describe accurate relaxation of initial disturbance with any wavelength.
This non-relativistic analysis gives suggestions for our next work of relativistic analysis of relaxation.
}

\begin{document}

\maketitle

\section{\label{sec:level1}INTRODUCTION}
Macroscopic phenomena of dissipative fluid are in general described by Navier-Stokes equations. 
However, Navier-Stokes equations can not provide good descriptions 
if the characteristic length of phenomenon is comparable to the mean free path of the constituent particles. 
For this reason Navier-Stokes equation cannot describe shock structure and short wavelength sound wave propagation
\cite{Meyer,Green,Schotter} that are important in astrophysics and plasma physics. 

To study short wavelength phenomena 
we have to use kinetic theory description, 
such as by the Boltzmann equation.
Unfortunately it is quite difficult to solve the Boltzmann equation and various methods for obtaining approximate solutions 
have been developed. 
The most standard method is Chapman-Enskog expansion~\cite{c-e} that was developed by Chapman, Enskog and Hilbert. 
This method begins with an expansion of distribution function in the
 mean free path of constituent particles and simplifies the results at a given order by introducing 
expressions from the previous order terms. 
This method derives the Euler equations at zeroth order and the Navier-Stokes equations at 
the first order and determines the transport coefficients that cannot be determined by Navier-Stokes equations. 
For this reason it is natural to expect that higher order approximations improve the description. 
The second and third order approximation was proposed by Burnett.\cite{burnett1,burnett2,Kim_Hayakawa(2003)} 
However these equations provide little or no improvement in describing short wavelength phenomena. 
It is because the Chapman-Enskog expansion is a sort of iteration process and therefore inappropriate for a series that is not convergent. 
Modification of the Chapman-Enskog expansion has been proposed, for example, by Chen $\textit{et al}$.\cite{chen1,chen3,chen2} 
who avoid the recycling of lower order results.
 This extension of the procedures provided improved descriptions of short wavelength phenomena in the case of BGK model of kinetic theory, 
though this procedure cannot describe phenomena whose Knudsen number is close to or larger than unity.

Another procedure is the moment method introduced by Grad.\cite{grad1,grad2}
This procedure is a expansion of the distribution function not in the mean free path but using Hermite polynomials 
and its 13-moment expansion reproduces the Navier-Stokes equation in the leading order. 
However this expansion is asymptotic rather than convergent so that the further development of expansion does not easily produce 
very accurate solutions in the short wavelength limit.\cite{reinecke,velasco,jou,muller}

In this paper, we derive the dispersion relation of BGK model~\cite{bgk} of the linearized kinetic equation 
and analyze the relaxation process using the solutions of the dispersion relation.
Since we study the kinetic model equation, 
the solution of the dispersion relation is correct in the domain of applicability of BGK model.
The accuracy of BGK model has been studied in many papers: 
the experimental data for phase velocities and attenuation rate~\cite{Meyer} indicates that BGK model is
almost accurate in any Knudsen number~\cite{bgk,st65} but inadequate for phenomena sensitive to high-order moments.\cite{gj}
We are interested in time evolution of initial disturbance 
so that we solve the dispersion relation in terms of the frequency and obtain all modes of relaxation. 
Since we use the solution of the dispersion relation of the kinetic equation, 
the solution accurately describes relaxation of initial disturbance with any wavelength 
and can be applied to various problem of the dilute gas dynamics; 
for example, applications include dilute gas dynamics around an object~\cite{Takata}, 
study of the ``atmospheres`` of cosmic bodies~\cite{Larina}, 
and the motion of a micrometer incident on the Earth's upper atmosphere~\cite{Coulson}.
%
In addition, 
this non-relativistic analysis may give suggestions to our next work of relativistic analysis of relaxation.

In Section \ref{sec:level2} we derive dispersion relations of BGK equation of the linearized kinetic equation. 
In Section \ref{sec:level7} we solve the dispersion relation obtained in Section \ref{sec:level2} 
and derive eigenfunctions for all modes. 
In Section \ref{sec:level10} we study the corresponding physical meaning of each eigenfunction. 


\section{\label{sec:level2}THE LINEARIZED KINETIC EQUATIONS AND THE DISPERSION RELATION}
In this section, we derive dispersion relations of BGK model for the linearized kinetic equation.\cite{bgk,cerbooknonrela}
The detailed calculation is presented in Appendix~\ref{sec:detailcal}.

When there is no external field, BGK model of kinetic equation is
\begin{align}
\frac{D f(t,{\bf x},{\bf v})}{Dt} &= - \frac{f(t,{\bf x},{\bf v})-f_{eq}(t,{\bf x},{\bf v})}{\tau}
\label{eq:bgk} 
,\\
\frac{D}{Dt}&=\left( \frac{\partial}{\partial t} + {\bf v} \cdot \nabla  \right)
.
\end{align}
In Eq.~(\ref{eq:bgk}), $\tau$ is the relaxation time and $f_{eq}$ represents the local Maxwell-Boltzmann distribution 
\begin{align}
f_{eq} (t,{\bf x},{\bf v}) &= \frac{\rho(t,{\bf x})}{m (2 \pi R T(t,{\bf x}))^{3/2}} e^{-\frac{({\bf v-u(t,{\bf x})})^2}{2RT(t,{\bf x})}}
,
\end{align}
where
\begin{align}
R = k_B / m,
\end{align}
and $k_B$ is the Boltzmann constant.

Eq.~(\ref{eq:bgk}) is a non-linear equation for $f(t,{\bf x,v})$ 
because of the non-linear dependence of $f_{eq}$ on $f$ through the following conditions:
\begin{align}
&\int \left( f_{eq} - f \right) \psi^{\mu} d^3 {\bf v} = 0
,
\\
&\psi^{\mu} = m \left(1,\: {\bf v},\: \frac{1}{2} (\bf{v - u})^2 \right)
.
\end{align}
These conditions are called matching conditions.

For obtaining the dispersion relation, we start by expanding the distribution function 
around an equilibrium state $f_0({\bf v})$
\begin{align}
\delta f = f-f_0, \quad \delta f_{eq} = f_{eq} - f_0
.
\end{align}
The linearized BGK model of the kinetic equation becomes the following form:
\begin{align}
\left( \frac{\partial}{\partial t} + {\bf v} \cdot \nabla  \right) \delta f  
= - \frac{\delta f - \delta f_{eq}}{\tau}
\label{eq:linbol}
.
\end{align}

We assume the following form of solution:
\begin{equation}
\delta f =\delta \tilde{f} e^{- i \omega(t-t_0)+i \mathbf{k \cdot x}} 
\label{eq:ome}
.
\end{equation}
Then, Eq.~(\ref{eq:linbol}) is
\begin{equation}
\left( \frac{1}{\tau} -i \omega + i {\bf k \cdot v} \right) \delta f = \frac{1}{\tau} \delta f_{eq}
,
\end{equation}
where
\begin{align}
\delta f_{eq} &= f_0 \left[ \frac{\delta \rho}{\rho_0} + \frac{{\bf v \cdot \delta u}}{R T_0} + \left(\frac{v^2}{2 R T_0} 
- \frac{3}{2} \right) \frac{\delta T}{T_0} \right] 
\label{eq:f0},
\\
f_0 &= \frac{\rho_0}{m (2 \pi R T_0)^{3/2}} e^{-v^2/2 R T_0}
\label{eq:f0b}
.
\end{align}

Using the matching conditions, we can rewrite $\delta \rho$,~$\delta {\bf u}$, and $\delta T$ as the integrals of $\delta f$
\begin{align}
\delta \rho(t,{\bf x}) &= \int m \delta f d^3 {\bf v}
,
\\
\delta {\bf u}(t,{\bf x}) &= \int \frac{m}{\rho_0} {\bf v} \delta f d^3 {\bf v}
,
\\
\delta T(t,{\bf x}) &= \int \frac{2 m}{3 \rho_0} \left(\frac{v^2}{2 R} - 
\frac{3}{2} T_0 \right) \delta f d^3 {\bf v}
.
\end{align}

Then Eq.~(\ref{eq:linbol}) becomes
\begin{align}
&\left( \frac{1}{\tau} -i \omega + i {\bf k \cdot v} \right) \delta f({\bf v}) 
\label{eq:mlbol} \\ \nonumber
&= \frac{f_0({\bf v})}{\tau} \int d^3 {\bf v'} 
\left[\frac{m}{\rho_0} + \frac{2 m}{\rho_0} \frac{{\bf v}}{\sqrt{2 R T_0}} \cdot \frac{{\bf v'}} 
{\sqrt{2 R T_0}} 
+ \frac{2 m}{3 \rho_0} \left(\frac{v^2}{2 R T_0} - \frac{3}{2} \right) \left( \frac{v'^2}{2 R T_0} - \frac{3}{2} \right)
\right] \delta f({\bf v'})
.
\end{align}

In the following we take $\tau$ as a unit of time, $\sqrt{2 R T_0}$ as a unit of velocity.
\begin{equation}
\omega \tau \equiv \bar {\omega}, \quad \sqrt{2 R T_0} \tau k \equiv \bar{k}, \quad \frac{v}{\sqrt{2 R T_0}} \equiv \bar{v}
.
\label{eq:diml}
\end{equation}
Hereafter we omit $\bar{}$.

Finally, the linearized equation of BGK model is
\begin{align}
\delta f({\bf v}) &= \int d^3 {\bf v'} K({\bf v,v'}) \delta f({\bf v'})
\label{eq:inteq}
,
\end{align}
where
\begin{align}
K({\bf v,v'}) \equiv \frac{m (2 R T_0)^{3/2}}{\rho_0} \frac{f_0({\bf v})}{1 -i \omega + i {\bf k \cdot v}}
\left[1 + 2 {\bf v \cdot v'} + \frac{2}{3} \left(v^2 - \frac{3}{2} \right) \left( v'^2 - \frac{3}{2}
\right) \right] 
.
\end{align}
Above equations make sense only when $1 - i \omega + i {\bf k \cdot v} \neq 0$.
We explain afterward the case where $1 - i \omega + i {\bf k \cdot v} = 0$.

Eq.~(\ref{eq:inteq}) is the homogeneous Fredholm integral equation of the second kind. In particular, the kernel function
$K({\bf v,v'})$ can be separated with respect to the variables ${\bf v}$ and ${\bf v'}$ and this equation can be solved according to the general 
procedure.

First we perform integration of Eq.~(\ref{eq:inteq}) with respect to ${\bf v}$. The equation becomes
\begin{align}
I_{11} \frac{\delta \rho}{\rho_0} + I_{12} \frac{\delta u_x}{\sqrt{2 R T_0}} + I_{13} \frac{\delta T}{T_0} = 0
,
\label{eq:cont}
\end{align}
where
\begin{align}
I_{11} &= \frac{1 - i \omega}{b} -\sqrt{\pi} e^{b^2} \mathrm{Erfc}(b)
,
\\
I_{12} &= 2 i \left( 1 - b \sqrt{\pi} e^{b^2} \mathrm{Erfc}(b) \right) 
,
\\
I_{13} &= - \left[ b - \left(b^2 + \frac{1}{2} \right) \sqrt{\pi} e^{b^2} \mathrm{Erfc}(b) \right]
,
\\
b &= \frac{1- i \omega}{k} 
\label{eq:b}
.
\end{align}

Secondly we multiply Eq.~(\ref{eq:inteq}) by ${\bf k \cdot v}$ and perform integration with respect to ${\bf v}$. The equation becomes
\begin{align}
I_{21} \frac{\delta \rho}{\rho_0} + I_{22} \frac{\delta u_x}{\sqrt{2 R T_0}} + I_{23} \frac{\delta T}{T_0} = 0
,
\label{eq:mom}
\end{align}
where
\begin{align}
I_{21} &= - i \left(1 - b \sqrt{\pi} e^{b^2} \mathrm{Erfc}(b) \right)
,
\\
I_{22} &= - k + 2 b \left(1 - b \sqrt{\pi} e^{b^2} \mathrm{Erfc}(b) \right)
,
\\
I_{23} &= i b \left\{b - \left(b^2 + \frac{1}{2} \right) \sqrt{\pi} e^{b^2} \mathrm{Erfc}(b) \right\}
.
\end{align}

Thirdly we multiply Eq.~(\ref{eq:inteq}) by $v_{\perp}$ and perform integration with respect to ${\bf v}$. The equation becomes
\begin{align}
&I_{\perp \perp} \frac{\delta u_{\perp}}{\sqrt{2 R T_0}} = 0
,
\\
&I_{\perp \perp} = k - \sqrt{\pi} e^{b^2} \mathrm{Erfc}(b)
.
\end{align}

Finally we multiply Eq.~(\ref{eq:inteq}) by $v^2$ and perform integration with respect to ${\bf v}$. The equation becomes
\begin{align}
I_{31} \frac{\delta \rho}{\rho_0} + I_{32} \frac{\delta u_x}{\sqrt{2 R T_0}} + I_{33} \frac{\delta T}{T_0} = 0
,
\label{eq:en}
\end{align}
where
\begin{align}
I_{31} &= \frac{1}{2} \left[3 k - 2 \left\{b - (b^2 - 1) \sqrt{\pi} e^{b^2} \mathrm{Erfc}(b) \right\} \right]
,
\\
I_{32} &= - \left[2 b \left\{b - (b^2 - 1) \sqrt{\pi} e^{b^2} \mathrm{Erfc}(b) \right\} - 3 \right]
,
\\
I_{33} &= \frac{1}{2} \left[3 k - \left\{2b (1 - b^2)+(2 b^4 - b^2 + 1) \sqrt{\pi} e^{b^2} \mathrm{Erfc}(b) \right\} \right]
.
\end{align}

If the determinant of the above homogeneous system is set equal to zero,
the following dispersion relation is obtained:
\begin{eqnarray}
 \begin{vmatrix}
   I_{11} & I_{12} & I_{13} & 0 & 0 
   \\
   I_{21} & I_{22} & I_{23} & 0 & 0 
   \\
   I_{31} & I_{32} & I_{33} & 0 & 0
   \\
   0 & 0 & 0 & I_{\perp \perp} & 0 
   \\
   0 & 0 & 0 & 0 & I_{\perp \perp} 
 \end{vmatrix}
= 0
.
\end{eqnarray}
This condition implies either
\begin{eqnarray}
 \begin{vmatrix}
   I_{11} & I_{12} & I_{13}
   \\
   I_{21} & I_{22} & I_{23}
   \\
   I_{31} & I_{32} & I_{33}
 \end{vmatrix}
= 0
\label{eq:fuldisp}
,  
\end{eqnarray}
or
\begin{eqnarray}
I_{\perp \perp} =0
\label{eq:shea}
.
\end{eqnarray}
%
%
%
Eq.~(\ref{eq:fuldisp}) corresponds to longitudinal mode ($\delta u_x \neq 0,~\delta u_{\perp} = 0$), and 
Eq.~(\ref{eq:shea}) corresponds to transverse mode ($\delta u_x = 0,~\delta u_{\perp} \neq 0$).

Finally we obtain the exact solution $\delta f$ in the following form:
\begin{align}
&\delta f({\bf v}) = \sum_n \frac{C_n f_0({\bf v})}{1 -i \omega_n + i {\bf k \cdot v}} 
\label{eq:genesol} 
\left[\delta \rho_{\omega_n} + \frac{{\bf v} \cdot \delta {\bf u}_{\omega_n}}{R T_0} + \left(\frac{v^2}{2 R T_0} - \frac{3}{2} \right) 
\delta T_{\omega_n} \right] e^{- i (\omega_n t + {\bf k \cdot x})} 
,
\end{align}
where $C_n$ is a constant coefficient and $\delta \rho_{\omega_n}$, $\delta \mathbf{u}_{\omega_n}$, $\delta T_{\omega_n}$ are eigenfunctions obtaining 
from Eq.~(\ref{eq:fuldisp}).

We should study the case where $1 - i \omega + i {\bf k \cdot v} = 0$. 
In this case the mode becomes continuous. 
According to Eq.~(\ref{eq:mlbol}) the eigenfunction for this mode satisfies the following equation:
\begin{align}
0 &= \int d^3 {\bf v'} \frac{f_0({\bf v})}{\tau} \left[\frac{m}{\rho_0} + \frac{m}{\rho_0} \frac{{\bf v}}{R T_0} \cdot {\bf v'} 
+ \frac{2 m}{3 \rho_0} \left(\frac{v^2}{2 R T_0} - \frac{3}{2} \right) \left( \frac{v'^2}{2 R T_0} - \frac{3}{2} \right)
\right] \delta f({\bf v'})
\label{eq:con0} \\ \nonumber
&= \delta f_{eq}
.
\end{align}
This mode represents decaying of the moments of $f$ with vanishing $\delta f_{eq}$, i.e., $\delta \rho = \delta T = 0$, ${\bf \delta u = 0}$. 


\section{\label{sec:level7}RESULT}

\subsection{\label{sec:level8}DISPERSION RELATION}
\begin{figure}[htb]
  \parbox{\halftext}{
  \includegraphics[width=6cm,clip]{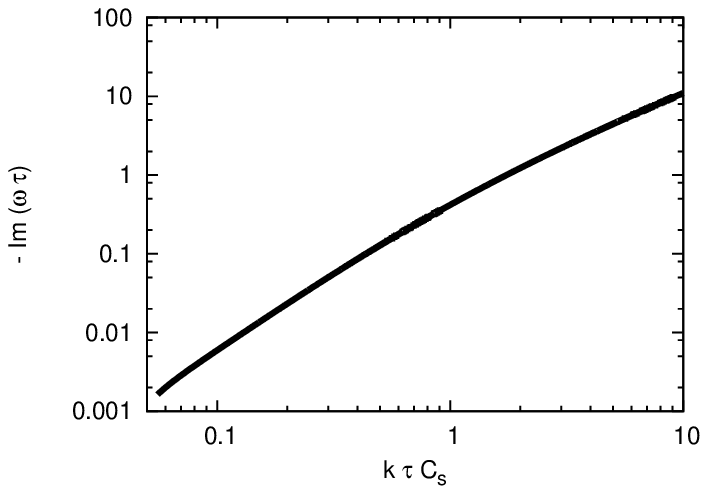}
     \caption{\label{fig:1}The decay rate of the thermal conduction mode.}}
  \hfill
  \parbox{\halftext}{
  \includegraphics[width=6cm,clip]{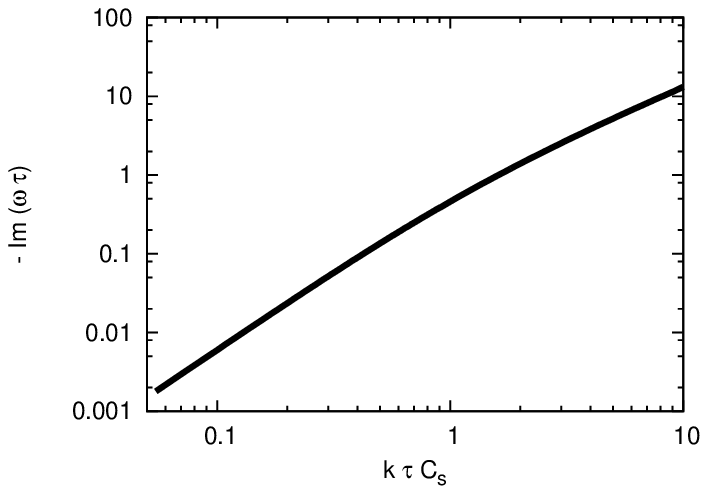}
     \caption{\label{fig:3}The decay rate of the shear flow mode.}}
\end{figure}

\begin{figure}[htb]
  \parbox{\halftext}{
  \includegraphics[width=6cm,clip]{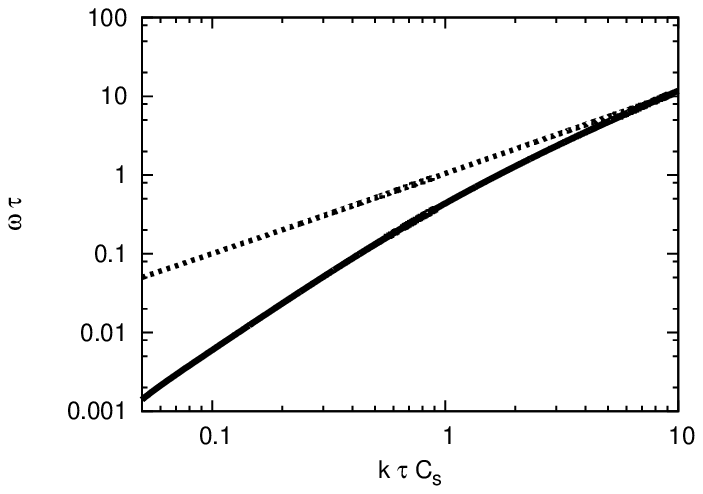}
     \caption{\label{fig:2}The dispersion relation of the sound wave mode.
     The Solid line represents the decay rate ($-$Im~$\omega$) 
     and the dotted line represents the frequency Re~$\omega$.}}
  \hfill
  \parbox{\halftext}{
  \includegraphics[width=6cm,clip]{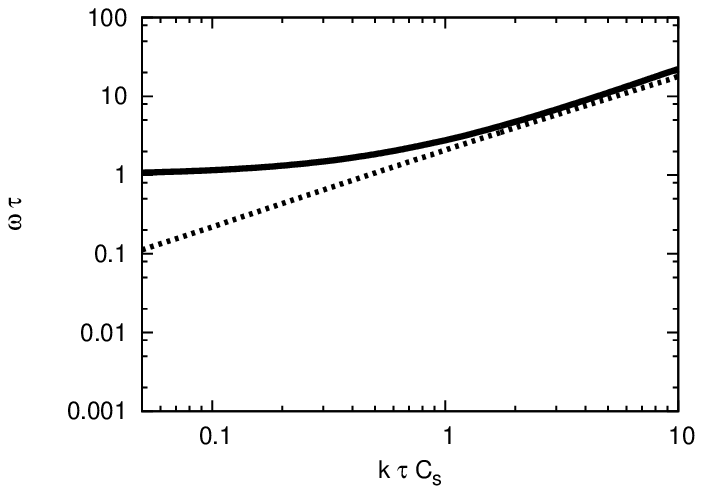}
  \caption{\label{fig:kin3}The dispersion relation of the transverse kinetic mode.
  The Solid line represents the decay rate ($-$Im~$\omega$) and
  the dotted line represents the frequency Re~$\omega$.}}
\end{figure}

In this section we show dispersion relations obtained in the previous section. First, we solve the dispersion relations numerically 
and results are shown in Fig.~\ref{fig:1}, Fig.~\ref{fig:3}, Fig.~\ref{fig:2}.


Solid lines represent the decay rate $(-\mathrm{Im}~\omega)$ and dotted lines represent the frequency $\mathrm{Re}~\omega$. 
In the long wavelength part, solid lines are proportional to $k^2$ and reproduce the result that can be obtained by
the first Chapman-Enskog approximation.
Regarding shear waves, we find only imaginary part of frequency and it indicates that shear waves cannot propagate in fluid when we 
consider initial value problems.

\begin{figure}[htb]
  \parbox{\halftext}{
  \includegraphics[width=6cm,clip]{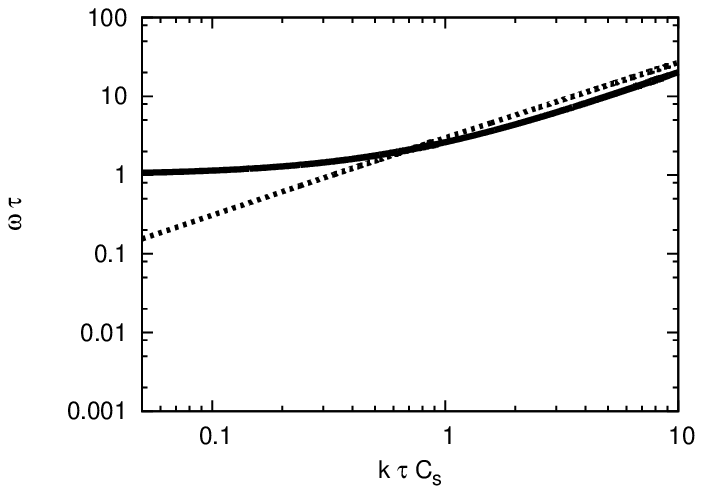}
   \caption{\label{fig:kin1}The dispersion relation of the longitudinal kinetic mode.
   The Solid line represents the decay rate ($-$Im~$\omega$) and
   the dotted line represents the frequency Re~$\omega$.}}
  \hfill
  \parbox{\halftext}{
  \includegraphics[width=6cm,clip]{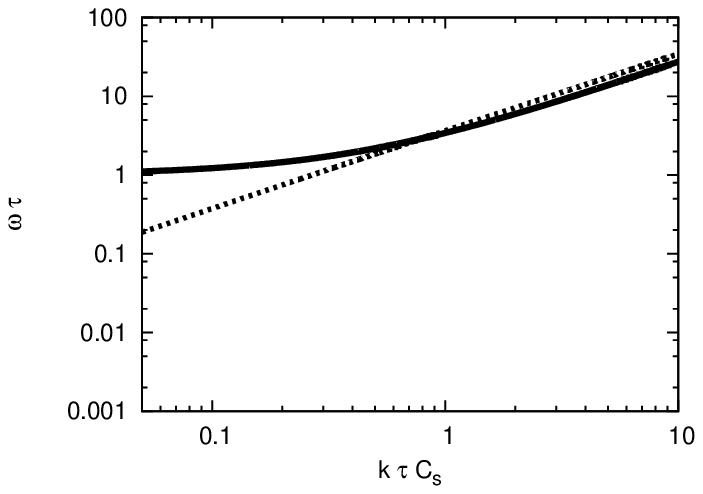}
     \caption{\label{fig:kin2}Another dispersion relation of the longitudinal kinetic mode.
     The Solid line represents the decay rate ($-$Im~$\omega$) and
     the dotted line represents the frequency Re~$\omega$.}}
\end{figure}

We also find rapidly decaying modes $- \mathrm{Im}~\omega \gtrsim 1 / \tau$ in Fig.~\ref{fig:kin1}, Fig.~\ref{fig:kin2}, 
and Fig.~\ref{fig:kin3}.
Since $(-\mathrm{Im}~\omega)$ is $1 / \tau$ at $k = 0$, 
these modes represent kinetic modes that do not appear in macroscopic descriptions.
We discuss kinetic modes in the following section.



\begin{figure}[htb]
  \parbox{\halftext}{
  \includegraphics[width=6cm,clip]{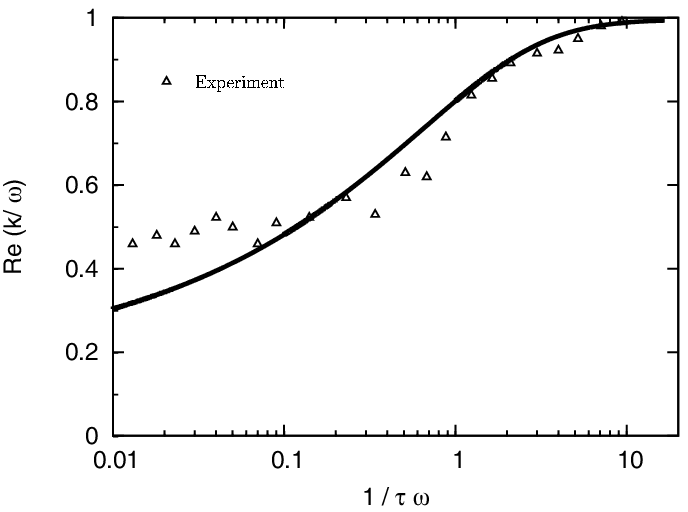}
     \caption{\label{fig:4}Phase velocity of sound wave mode 
              in units of the adiabatic sound speed, $\sqrt{\gamma R T_0}$.
              The dispersion relation is compared to the experiment of 
              Meyer and Sessler~\cite{Meyer}.}}
  \hfill
  \parbox{\halftext}{
  \includegraphics[width=6cm,clip]{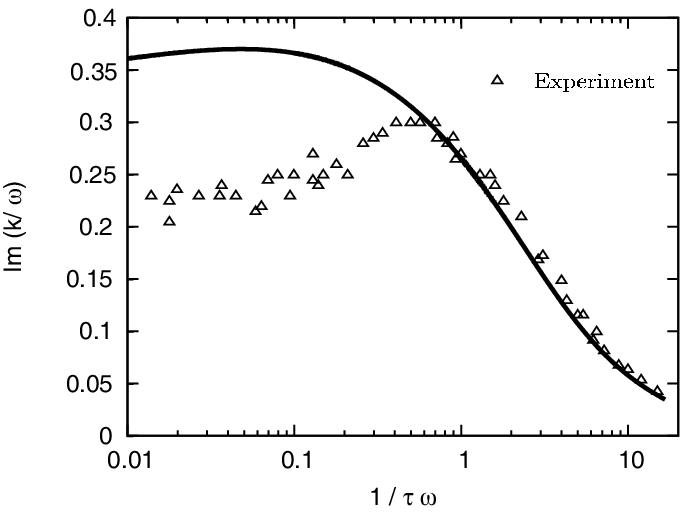}
     \caption{\label{fig:5}Decay rates of sound wave mode
              in units of the adiabatic sound speed, $\sqrt{\gamma R T_0}$.
              The dispersion relation is compared to the experiment of 
              Meyer and Sessler~\cite{Meyer}.}}

\end{figure}

Fig.~\ref{fig:4} and Fig.~\ref{fig:5} plot inverse of phase velocity and decay rates of sound wave modes 
that are obtained by solving Eq.~(\ref{eq:fuldisp}) for $k$ as a function of given $\omega$. 
These reproduce the results of Bhatnagar~\cite{gj} 
and are analogous to the results of Sirovich~\cite{st65} 
in which they have performed linear analysis of the Boltzmann equation for various forms of collision term.
We show only sound wave mode for $k$ but have found other modes, for example thermal wave modes.
%

\subsection{\label{sec:level9} EIGENFUNCTIONS}
In this section 
we show eigenfunctions to study physical picture of each mode.
 Substituting dispersion relations obtained in Section
\ref{sec:level8} for Eq.~(\ref{eq:cont}),~(\ref{eq:mom}),~(\ref{eq:en}), 
each mode is plotted with respect to the fluctuations 
$\delta \rho$, $\delta u_x$ and $\delta T$ that are normalized in $\delta \rho / \rho_0 + \delta u_x / \sqrt(2 R T_0) + \delta T / T_0 = 1$. 
\begin{figure}[htb]
  \parbox{\halftext}{
  \includegraphics[width=6cm,clip]{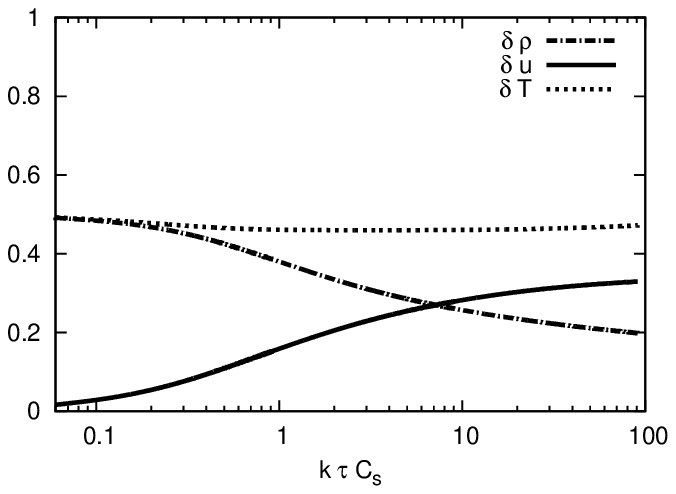}
     \caption{\label{fig:6}Eigenfunction of thermal conduction mode.}}
  \hfill
  \parbox{\halftext}{
  \includegraphics[width=6cm,clip]{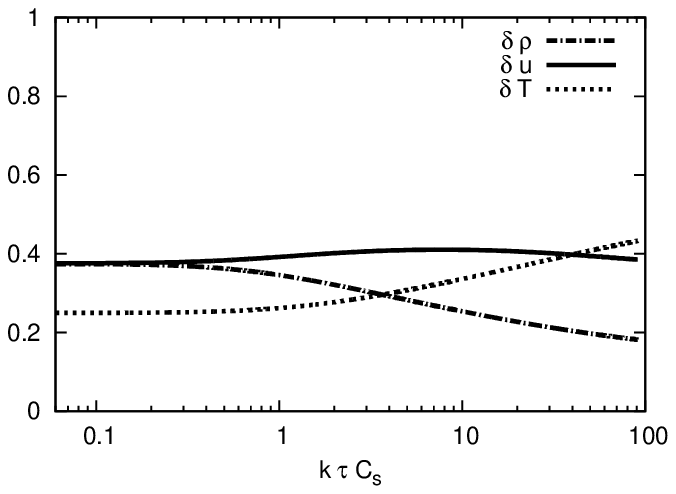}
     \caption{\label{fig:7}Eigenfunction of sound wave mode.}}
\end{figure}
\begin{figure}[htb]
  \parbox{\halftext}{
  \includegraphics[width=6cm,clip]{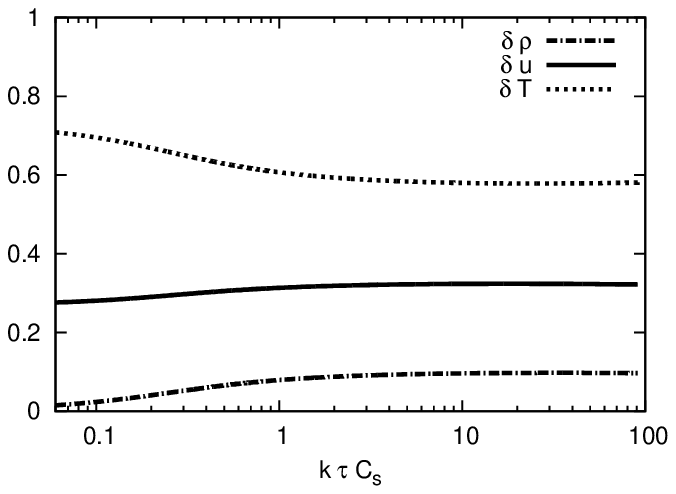}
     \caption{\label{fig:8}Eigenfunction of the longitudinal kinetic mode.}}
  \hfill
  \parbox{\halftext}{
  \includegraphics[width=6cm,clip]{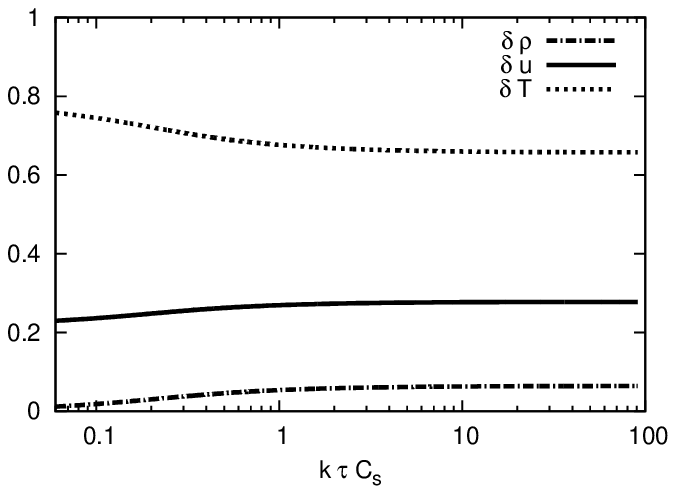}
     \caption{\label{fig:9}Eigenfunction of the second longitudinal kinetic mode.}}
\end{figure}


In Fig.~\ref{fig:6}, $\delta T$ is dominant and $\delta u_x$ is very small for a long wavelength part, 
which indicates that this mode represents thermal conduction. 
Interestingly, $\delta u_x$ increases with decreasing wavelength 
and it indicates that picture of pure thermal conduction cannot be applied to the short wavelength part of this mode. 

In Fig.~\ref{fig:7}, $\delta \rho$ and $\delta u_x$ are dominant for large wavelength 
and this mode can be regarded as sound wave mode. 
In the short wavelength regime $\delta T$ dominates the others 
and it represents that this mode is oscillation between $\delta T$ and $\delta u_x$ in this regime.

Fig.~\ref{fig:8} and Fig.~\ref{fig:9} indicate that these two modes show similar relaxation. 
Both modes contain small $\delta \rho$ and large $\delta T$ and $\delta u_x$ for any wavelengths. 
However, the timescale of decay of these modes is shorter than relaxation time $\tau$ 
and these modes cannot propagate and do not appear on macroscopic scale.


\section{\label{sec:level10}DISCUSSION}
\subsection{\label{sec:level11}KINETIC DECAY}
In general, relaxation is achieved by collisions so that decay rate $(-\mathrm{Im}~\omega)$ is not simply expected to be larger than $1/\tau$. 
However, Fig.~\ref{fig:2} and Fig.~\ref{fig:3} indicate that it is contrary. 
In this section we discuss why this happens.

In a short wavelength regime the relaxation rate $(-\mathrm{Im}~\omega)$ of this mode is larger than $1/\tau$ 
and it contradicts our understanding that relaxation is achieved through collision. 
In the BGK model adopted in this paper, however, 
$\tau$ is average relaxation time, namely $\tau$ is independent of particle velocity $v$. 
Thus, we should consider that physical system become collisionless gas when k is larger. 

For obtaining physical description of relaxation in the collisionless regime, 
we consider macroscopic momentum equation.
From kinetic theory, linearized momentum equation is

\begin{equation}
\rho_0 \frac{\partial}{\partial t} \delta {\bf u} = - \nabla \delta p - \nabla \overleftrightarrow{\mathbb{P}}
,
\end{equation}
where $\overleftrightarrow{\mathbb{P}}$ is traceless stress tensor.
We consider only longitudinal mode. 
We multiply Eq~(\ref{eq:mom}) by $\omega$ and use ideal gas law: $p = \rho R T$.
Then, Eq~(\ref{eq:mom}) can be cast into the following form:
\begin{align}
&- \omega {\bf k \cdot \delta u} + k^2 (\delta \rho + \delta T) 
\label{eq:eigmom}
+ \left[- k^2 (\delta \rho + \delta T) + \omega \delta \rho \left(- i + \frac{2 b}{k} \omega \right)
\right.
\\
\nonumber
&\times \left. \left(1 - b \sqrt{\pi} e^{b^2} \mathrm{Erfc}(b) 
+ i \omega b \left\{b - \left(b^2 + \frac{1}{2} \sqrt{\pi} e^{b^2} \mathrm{Erfc}(b) \right) \right\} \right) \delta T
 \right] = 0
.
\end{align}

\begin{figure}[htb]
  \parbox{\halftext}{
  \includegraphics[width=6cm,clip]{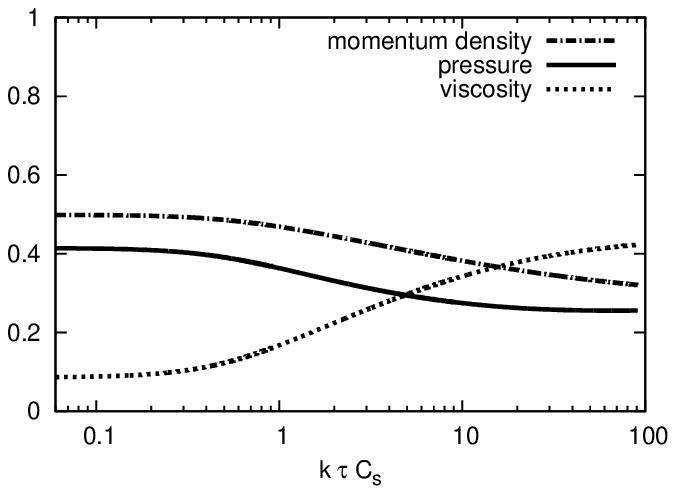}
     \caption{\label{fig:10}The relative magnitude of each term in the momentum equation for sound wave mode.}}
  \hfill
  \parbox{\halftext}{
  \includegraphics[width=6cm,clip]{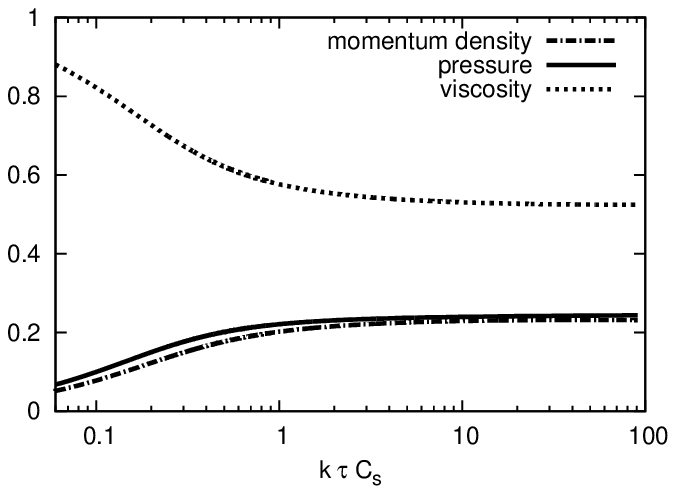}
     \caption{\label{fig:12}The relative magnitude of each term in the momentum equation for longitudinal kinetic mode.}}
\end{figure}

Fig.~\ref{fig:10} plots the ratio of magnitude of each term of the momentum equation in sound wave mode Eq~(\ref{eq:eigmom}).
This figure indicates that viscosity is small for a long wavelength and becomes larger for a short wavelength as in macro description.

%
Fig.~\ref{fig:12} is the same plot for kinetic modes.
These modes are purely kinetic and are not found in Chapman-Enskog approximation. 
Fig.~\ref{fig:12} indicates that viscosity dominates pressure and momentum density for a long wavelength. 
This is obviously different from the description of viscosity of Chapman-Enskog approximation 
in which viscosity is small correction to macroscopic variables.
This implies that these modes 
are purely kinetic modes.

\subsection{\label{sec:level101}ASYMPTOTIC ANALYSIS OF DISPERSION RELATION}
For studying the property of the dispersion relation, 
we take the long wavelength limit.
First we consider the shear flow mode. 
As in Sec.~\ref{sec:level2}, the dispersion relation of the shear flow is 
\begin{align}
\rho_0 \delta u_{\perp} &= m \int d^3 {\bf v} v_{\perp} \delta f
\nonumber
\\
&= m \int d^3 {\bf v} v_{\perp} \frac{f_{eq}}{1 - i \bar{\omega} + i \bar{k} \bar{v}_x} 
\left[ \frac{\delta \rho}{\rho_0} + \frac{{\bf v \cdot \delta u}}{R T_0} 
+ \left( \frac{v^2}{2 R T_0} - \frac{3}{2} \right) \frac{\delta T}{T_0} \right]
.
\end{align}
We expand the integrand in powers of $d = \bar{k} / (1 - i \bar{\omega})$ and 
neglect the terms higher than second order on the right-hand side. 
Then above equation reduces to
\begin{align}
\rho_0 \delta u_{\perp} &\simeq \int d^3 {\bf v} v_{\perp} \frac{1}{1 - i \bar{\omega}}
\left[ 1 - i d \bar{v}_x - d^2 \bar{v}_x^2  \right] \delta f_{eq}
.
\end{align}
Rewriting above equation, we get
\begin{align}
\rho_0 \delta u_{\perp} &= \frac{1}{1 - i \bar{\omega}} \int d^3 {\bf v} v_{\perp} 
( 1  - d^2 \bar{v}_x^2 ) \delta f_{eq}
\nonumber
\\ 
&= \frac{1}{1 - i \bar{\omega}}(1 - 2 \alpha_1 d^2) \rho_0 \delta u_{\perp}
.
\end{align}
where
\begin{equation}
\alpha_1 = \int d^3 {\bf v} \frac{m v_x^2 v_{\perp}^2}{\rho_0 (2 R T_0)^2} f_{eq}
.
\end{equation}
Neglecting $\bar{\omega}^2$, 
we obtain
\begin{equation}
\bar{\omega} = - 2 i \alpha_1 d^2 \simeq - 2 i \alpha_1 \bar{k}^2
.
\end{equation}

Next we study the long wavelength limit of thermal conduction mode 
and sound wave mode.
As in Sec.~\ref{sec:level2} the conservation of particle number is
\begin{align}
\delta \rho &= m \int d^3 {\bf v} \delta f
\nonumber
\\
&= m \int d^3 {\bf v} \frac{f_{eq}}{1 - i \bar{\omega} + i \bar{k} \bar{v}_x} 
\left[ \frac{\delta \rho}{\rho_0} + \frac{{\bf v \cdot \delta u}}{R T_0} 
+ \left( \frac{v^2}{2 R T_0} - \frac{3}{2} \right) \frac{\delta T}{T_0} \right]
.
\end{align}
The same as for the shear flow mode, 
we expand the integrand in powers of $d = \bar{k} / ( 1 - i \bar{\omega})$
and neglect the terms higher than second order on the right-hand side.
Then above equation reduces to
\begin{align}
\delta \rho &\simeq \frac{m}{1 - i \bar{\omega}} \int d^3 {\bf v} 
( 1 - i d \bar{v}_x - d^2 \bar{v}_x^2 ) \delta f_{eq}
\nonumber
\\
&=\frac{1}{1 - i \bar{\omega}} \left[ \delta \rho - i d \rho_0 \frac{\delta u_x}{\sqrt{2 R T_0}} 
- \frac{\rho_0 d^2}{2} \frac{\delta T}{T_0} \right]
,
\end{align}
where we use 
\begin{equation}
\int d^3 {\bf v} m v_x^2 \delta f_{eq} = \rho_0 R T_0 \left( \frac{\delta \rho}{\rho_0} + \frac{\delta T}{T_0} \right)
.
\end{equation}
Rewriting the above equation, we get
\begin{equation}
\left(i \bar{\omega} - \frac{d^2}{2} \right) \frac{\delta \rho}{\rho_0} 
- i d \frac{\delta u_x}{\sqrt{2 R T_0}} 
- \frac{d^2}{2} \frac{\delta T}{T_0} = 0
.
\end{equation}

Similarly, the conservation of the momentum and energy reduce to
\begin{align}
& - \frac{i d}{2} \frac{\delta \rho}{\rho_0}
+ (i \omega - 2 \alpha_4 d^2) \frac{\delta u_x}{\sqrt{2 R T_0}}
- \frac{i d}{2} \frac{\delta T}{T_0} = 0
,
\\
&\left[ \frac{3}{2} i \bar{\omega} - \alpha_2 d^2 \right] \frac{\delta \rho}{\rho_0}
- 2 i \alpha_2 d \frac{\delta u_x}{\sqrt{2 R T_0}}
+ \left[ \frac{3}{2} i \bar{\omega} - d^2 \left( \alpha_3 - \frac{3}{2} \alpha_2 \right) \right] \frac{\delta T}{T_0}
= 0
,
\end{align}
where
\begin{align}
\alpha_2 & = \int d^3 {\bf v} \frac{m v^2 v_x^2}{\rho_0 (2 R T_0)^2} f_{eq} = \frac{5}{4}
,
\\
\alpha_3 & = \int d^3 {\bf v} \frac{m v^4 v_x^2}{\rho_0 (2 R T_0)^3} f_{eq} = \frac{35}{8}
,
\\
\alpha_4 & = \int d^3 {\bf v} \frac{m v_x^4}{\rho_0 (2 R T_0)^2} f_{eq} = \frac{3}{4}
.
\end{align}


From above equations we obtain the dispersion relation in the following forms:
\begin{equation}
 - \frac{3 i}{2} \bar{\omega}^3 + \frac{19 d^2 \bar{\omega}^2}{4}
 + \left( \frac{5}{4} i d^2 + \frac{35}{8} i d^4 \right) \bar{\omega} - \frac{5 d^4}{8} - \frac{15 d^6}{16} = 0
.
\end{equation}
We obtain three roots given by $\bar{\omega_T}, \bar{\omega_{S \pm}}$ 
that are accurate to the second order in k
\begin{align}
\bar{\omega}_T &= - \frac{i d^2}{2} \simeq - \frac{i \bar{k}^2}{2}
\label{eq:thermalmode}
,
\\
\bar{\omega}_{S \pm} &= \pm \sqrt{\frac{5}{6}} d - \frac{4 i d^2}{3}
\label{eq:soundmode}
.
\end{align}
We substitute $d = \bar{k} / (1 - i \bar{\omega})$ into Eq.~(\ref{eq:soundmode}) 
and solve the equation with respect to $\bar{\omega}$. 
Then we obtain 
\begin{equation}
\bar{\omega}_{S \pm} \simeq \pm \sqrt{\frac{5}{6}} \bar{k} - \frac{i}{2} \bar{k}^2
\label{eq:smode}
.
\end{equation}


The above result shows that 
the decay rate (imaginary part) of the thermal conduction Eq.~(\ref{eq:thermalmode}) 
and the sound wave Eq.~(\ref{eq:smode}) is the same 
in the long wavelength limit.
It corresponds to the fact that the value of the Prandtl number is unity in the BGK model.
(The Prandtl number is the ratio of the viscosity times $C_p$ to the thermal conductivity.)



\subsection{\label{sec:level12}MATHEMATICAL STRUCTURE}
\subsubsection{\label{sec:level13}DISCRETE SPECTRUM}
In Sec.~\ref{sec:level2} we have derived the linearized equation of BGK model Eq.~(\ref{eq:inteq}). 
However, as is indicated at the last part of Sec.~\ref{sec:level2},
we assume $1 - i \omega + i {\bf k \cdot v} \neq 0$ when we derived Eq.~(\ref{eq:inteq}).

When we assume $1 - i \omega + i {\bf k \cdot v} \neq 0$, 
we can formally think the linearized equation of BGK model, Eq.~(\ref{eq:inteq}), as the following integral equation:
\begin{align}
\delta f({\bf v}) &= \lambda \int d^3 {\bf v'} K({\bf v,v'}) \delta f_{\lambda}({\bf v'})
\label{eq:lamgensol}
,
\end{align}
where
\begin{align}
\lambda = 1
, \\
K({\bf v,v'}) &\equiv \frac{m (2 R T_0)^{3/2}}{\rho_0} \frac{f_0({\bf v})}{1 -i \omega + i {\bf k \cdot v}}
\left[1 + 2 {\bf v \cdot v'} + \frac{2}{3} \left(v^2 - \frac{3}{2} \right) \left( v'^2 - \frac{3}{2} \right) \right] 
.
\end{align}
This is the homogeneous Fredholm equation of the second kind with eigenvalue $\lambda = 1$ 
and obtaining frequency $\omega$ is equivalent to deriving degenerated eigenfunction $\delta f_{\lambda \omega}$.

Mathematically, discrete modes that include both macroscopic modes and kinetic modes are degenerated solutions of integral equation 
Eq.~(\ref{eq:lamgensol}) and 
the number of discrete mode is equal to dimensions of eigenspace of eigenvalue $\lambda = 1$. 
In Eq.~(\ref{eq:lamgensol}), $K({\bf v,v'})$ is the compact integral operator. 
This indicates that the number of discrete modes is not infinite but finite because of the Fredholm alternative theorem.


\subsubsection{\label{sec:cont}CONTINUOUS SPECTRUM}
In Sec.~\ref{sec:level2} we obtained continuous spectrum $1 - i \omega + i {\bf k \cdot v} = 0$ in addition to the discrete modes. 
In this section we discuss the eigenfunction of this mode. 

We assume a situation that a beam of particles of constant velocity ${\bf v}_0$ is imposed in the initial condition. 
Then $\delta f$ becomes as follows:
\begin{align}
&\delta f = \delta f_1 + D \delta_D({\bf v - v_0})
,
\end{align}
where $D$ is constant coefficient and $\delta_D$ is the Dirac delta function. 
$D \delta_D({\bf v - v_0})$ represents particle flux distribution of ${\bf v = v_0}$. 
$\delta f_1$ represents the perturbed distribution function of gas. 

We substitute this $\delta f$ into Eq.~(\ref{eq:mlbol}) then it becomes
\begin{align}
& \left( \frac{1}{\tau} -i \omega + i {\bf k \cdot v} \right) \delta f_1 
\label{eq:cont11} 
+ \left( \frac{1}{\tau} -i \omega + i {\bf k \cdot v_0} \right) D
\\ \nonumber
&= \int d^3 {\bf v'} \frac{f_0({\bf v})}{\tau} \left[\frac{m}{\rho_0} + \frac{m}{\rho_0} \frac{{\bf v}}{R T_0} \cdot {\bf v'} 
+ \frac{2m}{3 \rho_0} \left(\frac{v^2}{2 R T_0} - \frac{3}{2} \right) \left( \frac{v'^2}{2 R T_0} - \frac{3}{2} \right)
 \right] \delta f_1({\bf v'})
\\ \nonumber
&+ \frac{f_0({\bf v})}{\tau} \left[\frac{m}{\rho_0} + \frac{m}{\rho_0} \frac{{\bf v}}{R T_0} \cdot {\bf v_0} 
+ \frac{2 m}{3 \rho_0} \left(\frac{v^2}{2 R T_0} - \frac{3}{2} \right) \left( \frac{v_0^2}{2 R T_0} - \frac{3}{2} \right)  \right] D
.
\end{align}

Then we assume next continuous spectrum.
\begin{align}
& \omega = - \frac{i}{\tau} + {\bf k \cdot v_0}
.
\end{align}

We substitute this $\omega$ into Eq.~(\ref{eq:cont11}). Then Eq.~(\ref{eq:cont11}) becomes
\begin{align}
\delta f_1 &= \int d^3 \frac{{\bf v'} f_0({\bf v})}{i \tau {\bf k} \cdot ({\bf v - v_0})} 
\left[\frac{m}{\rho_0} + \frac{m}{\rho_0} \frac{{\bf v}}{R T_0} \cdot {\bf v'} 
\label{eq:concon}
+ \frac{2 m}{3 \rho_0} \left(\frac{v^2}{2 R T_0} - \frac{3}{2} \right) \left( \frac{v'^2}{2 R T_0} - \frac{3}{2} \right)
 \right] \delta f_1({\bf v'})
\\ \nonumber
&+ \frac{f_0({\bf v})}{i \tau {\bf k} \cdot ({\bf v - v_0})} \left[\frac{m}{\rho_0} + 
\frac{m}{\rho_0} \frac{{\bf v}}{R T_0} \cdot {\bf v_0} 
+ \frac{2 m}{3 \rho_0} \left(\frac{v^2}{2 R T_0} - \frac{3}{2 T} \right) \left( \frac{v_0^2}{2 R T_0} - \frac{3}{2} \right)
 \right] D
.
\end{align}

The solution of this integral equation can be obtained the same as in Sec.~\ref{sec:level2}. We explain the case of shear flow for 
simplicity. We multiply Eq.~(\ref{eq:concon}) by $v_{\perp}$ and perform integration with ${\bf v}$. The equation becomes
\begin{align}
\left\{ k - \sqrt{\pi} e^{b^2} \mathrm{Erfc}(b) \right\} \delta u_{1y} = \sqrt{\pi} e^{b^2} \mathrm{Erfc}(b) \; D \: v_{0y}
,\\
b = - \frac{i {\bf k \cdot v_0}}{k} = - i v_{0x}
.
\end{align}

Finally we obtain eigenfunction $\delta u_{1y}$ in the following forms:
\begin{align}
& \delta u_{1y} = \left[\frac{k}{\sqrt{\pi} e^{- v_{0x}^2} \mathrm{Erfc}(- i v_{0x})} - 1\right]^{-1} D \: v_{0y} 
.
\end{align}
$\delta \rho_1$, $\delta {\bf u}_1$, $\delta T_1$ can be obtained with similar procedure.

\subsubsection{\label{sec:exactsol}GENERAL SOLUTION}
The continuous mode and the discrete mode we discussed in previous sections seem to cover all spectrum of Eq.~(\ref{eq:lamgensol}) 
and 
the general solution of the linearized BGK equation would be as follows:
\begin{align}
\delta f({\bf v}) &= \sum_n C_n K({\bf v,v'}) \delta f_{\omega_n}({\bf v'}) 
+ \int d^3 {\bf v_0} C({\bf v_0}) \hat{K}({\bf v,v'}) \delta f_{v_0}({\bf v'})
\label{eq:fredgensol}
, \\
\hat{K}({\bf v,v'}) &\equiv (1 - i \omega + i {\bf k \cdot v}) K({\bf v,v'})
.
\end{align}

Because of the Fredholm alternative theorem the number of the discrete modes is finite. 
We may conclude that most of the degrees of freedom of $\delta f$ belong to the continuous spectrum 
and only finite parts of them belong to the discrete modes.


\section{\label{sec:level14}CONCLUSION}
We have solved linearized kinetic equation of BGK model and obtain the exact solution for the relaxation of initial disturbances 
that describe the kinetic modes in addition to macroscopic modes that can be obtained by Chapman-Enskog expansion. 
Using those solutions, we have obtained the eigenfunction 
and analyzed the relaxation process corresponding to those solutions. 
Since we use the solution of the dispersion relation of the kinetic equation, 
the solution describes accurate relaxation of initial disturbance with any wavelength.
In next paper we will apply this method to the relativistic kinetic equation.


\section*{Acknowledgements}
We wish to thank Takayuki Muto and Takayuki Muranushi for fruitful discussions. We would also like to thank 
Professor Yoshio Tsutsumi for very helpful discussion about mathematical structure of linearized BGK model.

%

\appendix

\section{\label{sec:detailcal}THE DETAILED CALCULATION OF THE FLUID EQUATIONS}
In Sec.~\ref{sec:level2} we obtained BGK model of linearised kinetic equation in the following forms: 
\begin{align}
\delta f({\bf v}) &= \int d^3 {\bf v'} K({\bf v,v'}) \delta f({\bf v'}) 
,
\\
K({\bf v,v'}) &\equiv \frac{m (2 R T_0)^{3/2}}{\rho_0} \frac{f_0({\bf v})}{1 -i \omega + i {\bf k \cdot v}}
\left[1 + 2 {\bf v \cdot v'} + \frac{2}{3} \left(v^2 - \frac{3}{2} \right) \left( v'^2 - \frac{3}{2} \right) \right] 
,
\end{align}
where $f_0$ is the unperturbed state.

We study with the spherical coordinates in velocity space setting ${\bf k}$ as the z-axis. 
Then we have to relate ${\bf v \cdot \delta u}$ in $\delta f_{eq}$ to ${\bf k}$. 
We define $\theta$, $\theta'$, $\Theta$ as the angle measuring between ${\bf k}$ and ${\bf v}$, 
${\bf k}$ and $\delta {\bf u}$, 
${\bf v}$ and $\delta {\bf u}$, 
and $\phi$ as the angle between ${\bf v}$ and $\delta {\bf u}$ measuring on the plane projected perpendicular to ${\bf k}$. 
Then we have the following relation
\begin{align}
\cos \Theta = \cos \theta \cos \theta' + \sin \theta \sin \theta' \cos \phi \label{eq:kahou}
.
\end{align}

In this section we calculate the momentum equation of ${\bf k}$ direction as an example. 
We set x and y axis in the direction of ${\bf k}$ and $\delta {\bf u}_{\perp}$
 that is the projection of $\delta {\bf u}$ on the plane perpendicular to ${\bf k}$. Then $\delta u_x$ becomes
\begin{align}
k \delta u_x &= {\bf k} \cdot \int d^3 {\bf v} m {\bf v} \delta f
 \nonumber \\
&= 2 \pi m \int^{\infty}_{0} v^2 dv \int_{-1}^{1} d \mu \frac{v \mu \delta f_{eq}}
{1- i \omega + ikv \mu}
,
\end{align}
where $\mu=\cos \theta$. Then we replace $\delta f_{eq}$ with Eq.~(\ref{eq:f0}) 
and replace dimensional variables with dimensionless form as in Eq.~(\ref{eq:diml})
\begin{align}
\frac{2 k \rho_0}{\tau \sqrt{\pi}} & \int^{\infty}_{0} dv \int_{-1}^{1} d \mu \frac{\mu v^3 e^{-v^2}}{1- i \omega + ikv \mu} 
\left\{\left(v^2 - \frac{3}{2} \right) \delta T + \delta \rho  + 2 \frac{v \mu}{k} {\bf k \cdot \delta u} \right\}
.
\end{align}
We remove $\mu$ of odd index and the equation becomes
\begin{align}
\frac{2 k \rho_0}{\sqrt{\pi} \tau} & \int^{\infty}_{0} dv \int_{-1}^{1} d \mu \frac{\mu v^3}{k} \frac{e^{-v^2}}{b^2 + v^2 \mu^2}
\left\{- i v \mu \left(v^2 - \frac{3}{2} \right) \delta T - i v \mu \delta \rho + 2 b \frac{\mu v}{k} {\bf k \cdot \delta u} \right\}
\nonumber \\ 
 = \frac{4 \rho_0}{\sqrt{\pi} \tau} & \int^{\infty}_{0} dv \int_{0}^{1} d \mu  \frac{v^2 \mu^2 + b^2 - b^2}{b^2 + v^2 \mu^2} v^2 e^{-v^2}
\left\{- i \left(v^2 - \frac{3}{2} \right) \delta T - i \delta \rho + \frac{2 b}{k} {\bf k \cdot \delta u} \right\}
,
\end{align}
where 
\begin{align}
b =\frac{1-i \omega}{k}
.
\end{align}

Then we use next integral formula
\begin{align}
\int_{0}^{1} d \mu \frac{1}{a^2 + b^2 \mu^2} = \frac{1}{a b} \arctan \frac{b}{a}
,
\end{align}
and the equation can be integrated
\begin{align}
\frac{4 \rho_0}{\sqrt{\pi} \tau}& \int^{\infty}_{0} dv v^2 e^{-v^2} \left(1 - \frac{b}{v}\tan^{-1} \frac{v}{b} \right)
\left\{- i \left(v^2 - \frac{3}{2} \right) \delta T - i \delta \rho + \frac{2 b}{k} {\bf k \cdot \delta u} \right\}
\nonumber \\
=\frac{4 \rho_0}{\sqrt{\pi} \tau}& \int^{\infty}_{0} dv e^{-v^2} 
\left[ v^2 \left\{ - i \left(v^2 - \frac{3}{2} \right) \delta T - i \delta \rho + \frac{2 b}{k} {\bf k \cdot \delta u} \right\} \right.
\nonumber \\
&- \left. b \tan^{-1} \frac{v}{b} \left\{ i \left(- v^3 + v - v + \frac{3}{2} v \right) \delta T 
i v \delta \rho + \frac{2 b v}{k} {\bf k \cdot \delta u}  \right\} \right]
.
\end{align}

Using the identity
\begin{align}
\frac{d}{d v} (e^{- v^2} v^l) = (l v^{l-1} - 2 v^{l+1})e^{- v^2}
,
\end{align}
we perform partial integration
\begin{align}
\frac{4 \rho_0}{\sqrt{\pi} \tau}& \int^{\infty}_{0} dv \int_{0}^{1} d \mu e^{-v^2}
\left[ v^2 \left\{ - i \left(v^2 - \frac{3}{2} \right) \delta T - i \delta \rho + \frac{2 b}{k} {\bf k \cdot \delta u}  \right\} \right.
\nonumber \\
&\left. + \frac{b^2}{b^2 + v^2}
\left\{ i \left(\frac{v^2}{2} - \frac{1}{4} \right) \delta T + \frac{i}{2} \delta \rho 
- \frac{b}{k} {\bf k \cdot \delta u} \right\} \right]
.
\end{align}
We use next integral formulas
\begin{align}
\int^{\infty}_0 dx e^{-a^2 x^2} \frac{1}{x^2 +b^2} &= \frac{\pi}{2 b} e^{a^2 b^2} \mathrm{Erfc}(ab)
, \label{eq:error} 
\\
(a>0,~b>0)
 \nonumber \\
\mathrm{Erfc}(x) &= \frac{2}{\pi^{1/2}} \int^{\infty}_x e^{-t^2} dt , 
\\
\int^{\infty}_0 dx e^{-a^2 x^2} x^{2n} & = \frac{(2n-1)!!}{2^{n+1}} \sqrt{\frac{\pi}{a^{2n+1}}}
. \\
(a > 0) \nonumber
\end{align}
Finally the equation can be integrated as follows:
\begin{align}
\frac{\rho_0}{\tau} \left[i b \left\{b - \left(b^2 + \frac{1}{2} \right) \sqrt{\pi} e^{b^2} \mathrm{Erfc}(b) \right\} \delta T 
- i \left(\delta \rho + \frac{2 i b}{k} {\bf k \cdot \delta u} \right) \left(1 - b \sqrt{\pi} e^{b^2} \mathrm{Erfc}(b) \right) \right]
.
\end{align}

On the other hand, integration with respect to $\delta f_{eq}$ is clearly 
\begin{align}
{\bf k} \cdot \int d {\bf v} m {\bf v} \delta f_{eq} = \frac{\rho_0}{\tau} {\bf k \cdot \delta u}
.
\end{align}

In the end, the momentum equation is
\begin{align}
- {\bf k \cdot \delta u} + i b \left[b - \left(b^2 + \frac{1}{2} \right) \sqrt{\pi} e^{b^2} \mathrm{Erfc}(b) \right] \delta T 
\nonumber \\
- i \left(\delta \rho + \frac{2 i b}{k} {\bf k \cdot \delta u} \right) \left(1 - b \sqrt{\pi} e^{b^2} \mathrm{Erfc}(b) \right) = 0
.
\end{align}
The energy equation can be performed in the same way.

The shear flow equation can be calculated as follows. 
$\delta u_{\perp}$ is given by
\begin{align}
\rho_0 \delta u_{\perp} &= \int d^3 {\bf v} m v_{\perp} \delta f
\nonumber
\\
&= m \int d^3 {\bf v} v_{\perp} \frac{f_{eq}}{1 - i \omega + i k v_x}
\left[\left(v^2 - \frac{3}{2} \right) \delta
 T + \delta \rho 
+ 2 {\bf v \cdot \delta u} \right]
\nonumber
\\
&= m \int d^3 {\bf v} \frac{f_{eq}}{1 - i \omega + i k v_x} 2 v_{\perp}^2 \delta u_{\perp}
\label{eq:shear2}
.
\end{align}


Then after the same calculation of the momentum equation we obtain Eq.~(\ref{eq:shea}).

\section{ANALYTIC CONTINUATION}
Eq.~(\ref{eq:error}) can be used only when $b > 0$ and we have to regard b as $\vert b \vert$; 
if b has imaginary part, we have to replace $b$ to $-b$ in Eq.~(\ref{eq:error}). 
However in this paper we perform analytic continuation and remove these restrictions. 
In this section we explain the analytic continuation.

For simplicity we explain by using the transverse shear flow mode ($\delta u_{\perp} \ne 0$). 
In Eq.~(\ref{eq:shear2}) we consider the circular cylindrical coordinates $(v_{\perp},~\phi,~v_x)$ 
and perform $\phi$ integral.
Replacing to dimensionless form, the equation reduces to
\begin{align}
\int^{\infty}_{- \infty} d v_x  \int^{\infty}_{0} d v_{\perp} \frac{v_{\perp}^3}{i \{k v_x - i (\omega + i)\}}
 e^{-v_x^2-v_{\perp}^2}
.
\end{align}

Performing $v_{\perp}$ integral the equation becomes
\begin{align}
\frac{1}{2} \int^{\infty}_{- \infty} d v_x \frac{e^{-v_x^2}}{i \{k v_x - i (\omega + i)\}}
. \label{eq:an}
\end{align}

This integrand may have first-order pole on the real axis. 
However this singularity is removed to the upper half-plane if $\mathrm{Im}~\omega > -1$ 
and this results from the collision term of BGK model Eq.~(\ref{eq:bgk}). 
This fulfills the demand of the Landau method that this problem has to be considered as the Cauchy problem. 
For this reason the integration path has to be distorted below the singularity. 

Concerning above equation the restriction $\mathrm{Re}~b > 0$ is not satisfied when $\mathrm{Im}~\omega < -1$, 
so analytic continuation has to be performed. 
The equation has first-order pole so we use the principal integral
\begin{align}
\int^{\infty}_{-\infty} \frac{f(z) dz}{z - i 0} = P\int^{\infty}_{-\infty} \frac{f(z) dz}{z} + i \pi f(0)
.
\end{align}

The integration without analytic continuation corresponds to the integration path distorted above the singularity when $\mathrm{Re}~b<0$, 
so the principal integral is as follows:
\begin{align}
\frac{b}{2 k}\frac{\pi}{(-b)} e^{(-b)^2} \mathrm{Erfc}(-b) 
= \frac{1}{2} P \int^{\infty}_{- \infty} d v_{x} \frac{e^{-v_{x}^2}}{i(k v_{x} - i (\omega + i))} 
- \frac{i \pi}{2 i k}e^{-(ib)^2}
.
\end{align}

In the end, Eq.~(\ref{eq:an}) is as follows when $\mathrm{Re}~b<0$
\begin{align}
& \frac{1}{2} \int^{\infty}_{- \infty} d v_{x} \frac{e^{-v_{x}^2}}{i \{k v_{x} - i (\omega + i)\}}
\nonumber \\
&= \left(\frac{b}{2 k} \frac{\pi}{-b} e^{(-b)^2} \mathrm{Erfc}(-b) + \frac{\pi i}{2 i k} e^{-(i b)^2} \right) 
 + \frac{\pi i}{2 i k} e^{-(i b)^2}
\nonumber \\
&= \frac{\pi}{2 k} e^{b^2} (2-\mathrm{Erfc}(-b))
\nonumber \\
 &= \frac{\pi}{2 k} e^{b^2} \mathrm{Erfc}(b)
.
\end{align}
This indicates that analytic continuation permits us to use Eq.~(\ref{eq:shea}) even when $\mathrm{Re}~b<0$.


\begin{thebibliography}{99}
  
\bibitem{Meyer}
E. Meyer and G. Sessler, Z. Phys. {\bf 149}, (1957), 15-39.
\bibitem{Green}
M. Greenspan, J. Acoust. Soc. Am. {\bf 28}, (1956), 644-648.
\bibitem{Schotter} 
R. Schotter, Physics of Fluids, {\bf 17}, (1974), 1163.
\bibitem{c-e}
S. Chapman and T. G. Cowling, in \textit{The Mathematical Theory of Non-Uniform Gases} 3rd ed. (The University Press, Cambridge, 1991). 
\bibitem{burnett1}
D. Burnett, Proc. London Math. Soc. {\bf 39}, (1935), 385.
\bibitem{burnett2}
D. Burnett, Proc. London Math. Soc. {\bf 40}, (1935), 382.
\bibitem{Kim_Hayakawa(2003)}
H.-D .Kim \&  H. Hayakawa, Journal of the Physical Society of Japan, \textbf{72}, (2003), 1904.
\bibitem{chen1}
X. Chen, H. Rao and E. A. Spiegel, Phys. Rev. E, {\bf 64}, (2001), 046308.
\bibitem{chen3} 
X. Chen, H. Rao and E. A. Spiegel, Phys. Rev. E, {\bf 64}, (2001), 046309.
\bibitem{chen2}
E.~A. Spiegel and  J.-L. Thiffeault, Physics of Fluids, \textbf{15}, (2003), 3558.
\bibitem{bgk}
P.~L.~Bhatnagar, E.~P.~Gross, and M.~Krook Phys. Rev. {\bf 94}, 3, (1954), 511.
\bibitem{grad1}
H. Grad, in \textit{Thermodynamics of Gases}, Handbuch der Physik Vol. 12 (Springer-Verlag, Berlin, 1958).
\bibitem{grad2}
H. Grad, Commun. Pure Appl. Math. {\bf 2}, (1949), 331.
\bibitem{reinecke}
S. Reinecke and G. M. Kremer, Phys. Rev. A {\bf 42}, (1990), 815.
\bibitem{velasco}
R. M. Velasco and L. S. Garcia Colin, Phys. Rev. A {\bf 44}, (1991), 4961.
\bibitem{jou}
D. Jou, J. Casas-Vasquez, and G. Lebon, \textit{Extended Irreversible Thermodynamics}, (Springer, Heidelberg, 1993).
\bibitem{muller}
I. M$\ddot{\mathrm{u}}$ller and T. Ruggeri, \textit{Rational Extended Thermodynamics}, of Springer Tracts in Natural Philosophy, Vol. 37, 2nd ed. 
(Springer-Verlag, New York, 1998).
\bibitem{gj}
E. P. Gross and E. A. Jackson, Phys. Fluids {\bf 2}, (1959), 432-441.
\bibitem{Takata}
S. Takata, Y. Sone and K. Aoki, Physics of Fluids, {\bf 5}, (1993), 716.
\bibitem{Larina}
I. N. Larina, Fluid Dynamics, {\bf 17}, 5, (1982), 809.
\bibitem{Coulson}
S.~G. Coulson, Mon. Not. R. Astron. Soc. {\bf 332}, (2002), 741.
\bibitem{st65}
L. Sirovich and J. K. Thurber, J. Acoust. Soc. Am. {\bf 37}, (1965), 329-339.
\bibitem{st631}
L. Sirovich and J. K. Thurber, Phys. Fluids {\bf 6}, (1963), 10-20.
\bibitem{st632}
L. Sirovich and J. K. Thurber, Phys. Fluids {\bf 6}, (1963), 218-223.
\bibitem{book}
C. S. Wang Chang and G. E. Uhlenbeck, in \textit{STUDIES IN STATISTICAL MECHANICS V}, edited by J. De Boer and G. E. Uhlenbeck,
(North Holland, Holland, 1970).
\bibitem{cerbooknonrela}
C.~Cercignani, Theory and Application of the Boltzmann Equation, Scottish Academic Press, Edinburgh (1975).

\end{thebibliography}
\end{document}